# Non-reciprocal gain in non-Hermitian time-Floquet systems

Theodoros T. Koutserimpas and Romain Fleury*

*Laboratory of Wave Engineering, EPFL, 1015 Lausanne, Switzerland*

*To whom correspondence should be addressed. Email: romain.fleury@epfl.ch*

We explore the unconventional wave scattering properties of non-Hermitian systems in which amplification or damping are induced by time-periodic modulation. These non-Hermitian time-Floquet systems are capable of non-reciprocal operations in the frequency domain, which can be exploited to induce novel physical phenomena such as unidirectional wave amplification and perfect non-reciprocal response with zero or even negative insertion losses. This unique behavior is obtained by imparting a specific low-frequency time-periodic modulation to the coupling between lossless resonators, promoting only upward frequency conversion, and leading to non-reciprocal parametric gain. We provide a full-wave demonstration of our findings in a one-way microwave amplifier, and establish the potential of non-Hermitian time-Floquet devices for insertion-loss free microwave isolation and unidirectional parametric amplification.

In conventional media, wave scattering is usually reciprocal: the transmission coefficient from a source to a receiver remains the same if we interchange their locations or, in other words, the scattering matrix is always symmetric. This fundamental property of wave propagation is linked to microscopic reversibility, and holds for any linear time-invariant system in the absence of external time-odd bias [1,2]. There exist, however, many situations in which one would like to transmit waves unidirectionally by breaking reciprocity. For instance, non-reciprocal antennas that can emit and receive at the same frequency may allow for doubling the available bandwidth in the next generation of full-duplex telecommunication systems [3,4].

In electromagnetic systems, the conventional way of breaking Lorentz reciprocity is the use of magnetic materials and an external magnetic field as time-odd external bias [5]. Magnets, however, are bulky, expensive, and to a large extent incompatible with integrated circuit technology and standard CMOS fabrication methods. For this reason, magnet-free non-reciprocal components have been developed based on non-linearity [6-18] or time-modulation [19-32]. Non-linear systems strongly break reciprocity, however they are fundamentally limited as isolators [33]. Time-Floquet systems represent a promising alternative to magnetic-free isolation, but they do not conserve frequency, and are thus associated with unavoidable insertion losses due to the energy lost in all the intermodulation frequency channels.

In a different field of research, non-Hermitian photonic systems that exploit the interplay between gain, loss, and the coupling between individual optical components have created a wealth of new opportunities in classical physics to generate and control the transmission of light [34-46]. While non-Hermitian systems by themselves cannot break reciprocity [47], they provide an interesting platform for non-unitary scattering, loss compensation, and wave amplification [39-46]. For instance, PT symmetric systems can exhibit resonant localized amplification when operated in the broken phase, a phenomenon that can be exploited to obtain non-linear responses at much lower power than in the exact phase [44], and induce low-threshold optical non-reciprocity [45,46].

In this Letter, we investigate the unexplored physics of non-Hermitian systems for which non-Hermiticity is *not* due to the direct presence of material losses or gain, but is instead *induced by periodic time-modulation*. We show that these non-Hermitian time-Floquet systems can be obtained by considering time-dependent coupling between lossless components, leading to a unique way to generate parametric loss or gain. Simultaneously, the breaking of time-invariance in these systems can trigger non-reciprocal frequency conversions, allowing for highly efficient non-reciprocal amplification of the signal. By engineering at the same time non-Hermiticity and non-reciprocity, we overcome the insertion loss challenges inherent to Hermitian time-Floquet isolators, and obtain large non-reciprocal isolation with zero, or even negative insertion losses (signal amplification). We provide a full-wave demonstration of these findings at microwave frequencies.

Let us consider a simple time-invariant system described by the general two-by-two Hamiltonian $\mathbf{H} = [\omega_1, k_{12}; k_{21}, \omega_2]$. Due to microscopic reversibility, we must have $k_{21} = k_{12}^*$, which implies that we cannot create a non-Hermitian matrix through the off-diagonal components. Therefore, to induce non-Hermiticity, the only possibility is to make $\omega_1$ and $\omega_2$ complex. This is the usual solution employed in the literature, which corresponds to adding gain and/or loss to the medium inside the resonators. However, imagine that the coefficients $k_{12}$ and $k_{21}$ depend on time: microscopic reversibility now implies $k_{21}(t) = k_{12}^*(-t)$, meaning that the system can potentially be non-Hermitian ($k_{21}(t) \neq k_{12}^*(t)$) when $k_{12}(t)$ is not an even function of time. A particular example of such non-Hermitian systems is obtained for $k_{12}(t) = K(t) = k_0 + \Delta k' \cos(\Omega t) + j \Delta k'' \sin(\Omega t)$, with $j^2 = -1$, for which the relation $k_{21}(t) = k_{12}^*(-t)$ implies $k_{21}(t) = k_{12}(t) =$

$K(t)$, i.e. a non-Hermitian Hamiltonian with identical complex off-diagonal terms. In this letter, we focus on the class of two-level systems for which $\mathbf{H} = [\omega_1, K(t); K(t), \omega_2]$. Such Hamiltonians are not only non-Hermitian, but periodic in time with period $T = 2\pi/\Omega$, corresponding to time-Floquet non-Hermitian systems. Note that $k_0$ represents the static Hermitian part of the coupling, $\Delta k' \cos(\Omega t)$ the time-dependent Hermitian part, and $j\Delta k'' \sin(\Omega t)$ the time-dependent non-Hermitian part. In the following, we study the effect of each part of the coupling on the system dynamics and scattering properties.

An example of device described by such Hamiltonian is represented in Fig. 1. Two coupled resonators (coupling $K(t)$), each with different resonance frequency $\omega_1$ and $\omega_2$ ($\omega_1 < \omega_2$), are combined with two waveguides functioning as ports for incoming or outgoing wave signals, with decay rates $\gamma_1$ and $\gamma_2$. After applying Floquet theorem [48] in time domain, we use standard temporal coupled mode theory (CMT) [49,50], writing the amplitudes $\alpha_1$ and $\alpha_2$ inside the two resonators as $\alpha_{1,2}(t) = \sum a_{1,2}^n e^{j(\omega+n\Omega)t}$, where $a_1^n$ and $a_2^n$ represent the time-independent complex amplitudes of the n-th Floquet harmonic, and $\omega$ is the excitation frequency. Assuming excitation from port 1, the coupled-mode equations of the system are found as [51]

$$(\omega + n\Omega - \omega_1 - j\gamma_1)a_1^n - k_0 a_2^n - \left(\frac{\Delta k' + \Delta k''}{2}\right)a_2^{n-1} - \left(\frac{\Delta k' - \Delta k''}{2}\right)a_2^{n+1} = \sqrt{2\gamma_1}\delta_{n0} \qquad (1)$$

$$(\omega + n\Omega - \omega_2 - j\gamma_2)a_2^n - k_0 a_1^n - \left(\frac{\Delta k' + \Delta k''}{2}\right)a_1^{n-1} - \left(\frac{\Delta k' - \Delta k''}{2}\right)a_1^{n+1} = 0 \qquad (2)$$

Looking at the above equations, we see that the effect of $k_0$ is to couple identical frequencies, whereas the temporal modulation generates infinite but discrete frequency harmonics at $\omega + n\Omega$ for $n = \{-\infty, \ldots, -1, 0, +1, \ldots, +\infty\}$. More specifically, the harmonic time-Floquet modulation couples each resonator with the two adjacent Floquet harmonics in the other resonator. For a Hermitian system ($\Delta k'' = 0$), upward and downward frequency conversion are equally efficient. Very differently, adding a non-Hermitian time-Floquet modulation ($\Delta k'' \neq 0$) allows one to tune the amount of energy that will undergo upward and downward frequency transitions. Crucially, the condition $\Delta k' = \Delta k''$ is special as it enables only upward frequency conversion [51]. This unique capability is enabled by the non-Hermitian time-Floquet modulation.

To illustrate better the profound implication of this special condition on the scattering properties of the system, let us assume incidence from port 1 at $\omega = \omega_1$. In addition, from now on we fix the modulation frequency to $\Omega = \omega_2 - \omega_1$. Thus, the $n = +1$ harmonic is exactly at $\omega_2$ and will resonantly excite the second resonator. Fig. 2 shows the norm of the field amplitudes $a_{1,2}^n$ of the $n = -1$, $n = 0$ and $n = +1$ Floquet harmonics of in various cases of interest. In the Hermitian case (Fig. 2a, $\Delta k'' = 0$), the -1, 0 and +1 harmonics of the field amplitudes $a_{1,2}$ all have significant energy. When we increase $\Delta k'$, the energy of the -1 and +1 modes both increase, since all the frequency conversion rates are equal and proportional to $\Delta k'$. Still for $\Delta k'' = 0$, another interesting phenomenon occurs if we force $k_0$ to become zero (Fig. 2b): the system is driven to a condition where the second resonator cannot have a 0 harmonic, and transmission to the second port can only be at a frequency different than $\omega_1$. Yet, in both Hermitian cases of Fig.2a-b, because upward and downward frequency conversion coefficients are equal, the transmission coefficient for incidence at port 2 with frequency $\omega_2$ to port 1 with frequency $\omega_1$ would be the same than the transmission coefficient for an input field at port 1 at $\omega_1$ to port 2 at $\omega_2$, and the system is reciprocal.

The situation is drastically different when we provide the system with a non-Hermitian time-periodic modulation (Fig. 2c,d). The plots are made under the special condition $\Delta k' = \Delta k''$, which promotes absolute upward frequency conversion. As discussed above, and consistent with the CMT equations (1) and (2), this modulation never excites the $-1$ harmonic, whose amplitude is always identically zero, and forces the system to exhibit only the 0 and +1 Floquet harmonics (Fig.2c). In addition, if we force $k_0$ to become zero, the only remaining states are $a_1^0$ and $a_2^+$ (Fig. 2d). This means that the field is transmitted to port 2 only through upward frequency conversion. Under these conditions, by increasing $\Delta k' = \Delta k''$ we observe a linear increase of the amplitude of the upward-converted $a_2^+$ mode. Evidently, in these conditions a backward field incident on port 2 at $\omega_2$ will neither be downward converted to $\omega_1$ (since downward conversions are forbidden), nor be transmitted at $\omega_2$ through the zero harmonic (since $k_0 = 0$), and the transmission to port 1 will be *identically zero*. Therefore, the non-Hermitian time-Floquet system can exhibit perfect non-reciprocity: it transmits energy incident on port 1 to port 2, but has zero transmission for any signal incident on port 2. In [51], we provide a more complete study of the system's dynamics as $k_0$ and $\Delta k''$ change gradually, and demonstrate that this behavior remains true in a large vicinity of the condition $k_0 = \Delta k' - \Delta k'' = 0$.

Next, we move on to a quantitative study of the unique scattering properties of the Floquet-system. Still assuming time-harmonic incidence on port 1, with $\psi_{inc} = \psi_{inc}^0 e^{j\omega t}$, the transmitted field at the second port can be expanded as $\psi_t = \sum_n \psi_t^n e^{j(\omega + n\Omega)t}$, and we can define a transmission coefficient for each frequency harmonic as $S_{21}^n = \psi_t^n / \psi_{inc}^0$. Here, only the $n = -1,0,1$ harmonics are important and we use the short-hand notations $S_{21}^{n=-1} = S_{21}^-$, $S_{21}^{n=0} = S_{21}^0$ and $S_{21}^{n=+1} = S_{21}^+$. Similarly, we define transmission coefficients for excitation from port 2 and note them $S_{12}^-$, $S_{12}^0$ and $S_{12}^+$. The definition of scattering parameters are adequate since the system is faithfully linear, consistent with Eq. (1-2) (see [51] for an explicit demonstration). Note also that these quantities depend on the incident frequency, for instance $S_{21}^+(\omega)$ corresponds to the field transmitted at port 2 at $\omega + \Omega$ when $\omega$ is sent at port 1.

We represent in Fig. 3 the evolution of $|S_{21}^+(\omega_1)|$ and $|S_{12}^-(\omega_2)|$ as $\Delta k'$ is gradually increased, comparing excitation from port 1 at $\omega_1$, and the reciprocal situation of excitation from port 2 at $\omega_2$. We also look at the transmission at the incident frequency $|S_{21}^0(\omega_1)|$ and $|S_{12}^0(\omega_2)|$. We consider two cases: (i) the static Hermitian case with $k_0 = \Delta k'' = 0$ (Fig. 3a), and (ii) the non-Hermitian case with $k_0 = 0$, $\Delta k' = \Delta k''$ (Fig. 3b). Because $k_0 = 0$ in both cases, transmission at the excitation frequency is impossible, and we always find that the transmission coefficients at the incident frequency to the other side of the resonator ($S_{21}^0$ and $S_{12}^0$) are zero. In the Hermitian case, frequency conversion is symmetric and therefore the system is always reciprocal: we have $|S_{21}^+(\omega_1)| = |S_{12}^-(\omega_2)|$ regardless of $\Delta k'$. Conversely, in the non-Hermitian case, transmission from port 2 to port 1 is identically zero regardless of the considered harmonic, whereas $|S_{21}^+(\omega_1)|$ is non-zero, demonstrating the large non-reciprocal behavior. In addition, for sufficiently high values of $\Delta k'/\sqrt{\gamma_1 \gamma_2}$, we find that $|S_{21}^+(\omega_1)|$ can even reach values well above unity. Remarkably, the non-Hermitian time-Floquet modulation provides unidirectional parametric gain to the signal. This is also confirmed by direct Finite-Difference-Time-Domain (FDTD) simulations [51].

A closed-form analytical validation of this amplifying behavior can be obtained if we truncate the system of Eq. (1,2) to three Floquet harmonics. For a field incident at port 1 at frequency $\omega_1$, we find

$$S_{21}^+ = \frac{2\sqrt{\gamma_1\gamma_2}\Delta k'\left((-\Omega-j\gamma_2)(\Omega-j\gamma_1)+k_0^2\right)}{\left((-\Omega-j\gamma_2)(-j\gamma_1)-k_0^2\right)\left((-j\gamma_2)(\Omega-j\gamma_1)-k_0^2\right)} \quad (3)$$

If $k_0 = 0$, Eq. (3) becomes $S_{21}^+ = -2\Delta k'/\sqrt{\gamma_1\gamma_2}$ which is identical with the results in Fig. 3b, obtained numerically considering 201 Floquet harmonics. It is remarkable that amplification can be obtained with arbitrarily small modulation depths $\Delta k' = \Delta k''$ as long as the system is resonant enough, i.e. $\sqrt{\gamma_1\gamma_2} < \Delta k'$. On the other hand, the transmission coefficients of the 0 Floquet harmonic under the same conditions are dependent on $k_0$ and are given by:

$$S_{21}^0 = \frac{2k_0\sqrt{\gamma_1\gamma_2}}{(-j\gamma_1)(\Omega-j\gamma_2)-k_0^2}, \quad (4)$$

$$S_{12}^0 = \frac{2k_0\sqrt{\gamma_1\gamma_2}}{(\Omega-j\gamma_1)(-j\gamma_2)-k_0^2} \quad (5)$$

Clearly, these coefficients become zero when $k_0 = 0$, in perfect agreement with the results of Fig. 3.

We demonstrate our findings in a realistic full-wave scenario at microwave frequencies. The device has two split ring resonators with eigenfrequencies $f_1 = 3.1201$ GHz and $f_2 = 3.6921$ GHz, the two ports are microstrip transmission lines of width 0.333 mm and height 0.1778 mm, and the substrate is FR4 with $\varepsilon_r = 4.5$. Between the split ring resonators, we insert a capacitor of $C = 0.15849$ fF, which has a small circular modulation with depth $\Delta C = 15.849$ fF and slow modulation frequency $f_C = 572$ MHz, in series with a resistor modulated with $\Delta R = 1/\Delta C\omega$. We provide more details about the circuit implementation of these modulated elements in the supplementary material [51]. Note that in practical scenarios, we cannot completely eliminate the coupling strength $k_0$, however we can make it as small as we want. Thus, to minimize the constant coupling effect, we keep some distance between the components and place the resonators in antisymmetric positions (as shown in Fig. 4). The fields are computed using the three-dimensional Finite-Element Method (FEM) method in frequency domain using a harmonic balance simulation with a truncation of Maxwell's equations to three Floquet harmonics $\{-1, 0, +1\}$.

Fig. 4a shows the numerical results obtained from the full-wave FEM simulations (solid lines), and compares them to the predictions of our analytical CMT model (dashed lines). In Fig. 4a, we plot the spectrum in dB of the only transmission coefficients that are found to be non-zero: $|S_{21}^+(\omega)|$, and $|S_{21}^0(\omega)| = |S_{12}^0(\omega)|$. Note that the x-axis of Fig. 4a corresponds to the frequency *of the incident field*. A point of abscissa $\omega$ on the curve $|S_{21}^+(\omega)|$, for instance, describes the amplitude of the Fourier component at $\omega + \Omega$ of the field transmitted to the second port. Therefore, the peak of $|S_{21}^+|$ at the position $f_1$ of the x-axis corresponds to strong transmission to port two at $f_1 + \Omega/2\pi = f_2$, i.e. transmission via upward frequency conversion. The level of this peak is above 30 dB, meaning amplification. At any other frequency, including at $f_2$, the transmission from port 2 to port 1 $|S_{12}^0(\omega)|$ is always below -50 dB, i.e. $|S_{21}^+(f_1)|$ is more than 80dB higher than $|S_{12}^0(\omega)|$ over the entire spectrum, which demonstrates that this parametric amplification phenomenon is indeed strongly non-reciprocal and well-suited for signal isolation. In addition, we see that the analytical model captures very well the physics involved, the only discrepancies being attributed to the inherent dispersion of the coupling coefficients, which is neglected in CMT. We stress that these discrepancies are extremely small, as they correspond to corrections that are -100 dB below the incident field level. We find that the unidirectional transmission gain is linearly controlled by the modulation depth $\Delta C$ and the quality factor of the system, in perfect agreement with coupled mode theory. The simple Hamiltonian model is therefore a very good description of fully realistic systems.

Below the spectrum, we show the spatial distribution of the real part of the complex electric field (vertical component). Since this field is principally made up of two frequency components, one at $\omega_1$ and another at $\omega_2$, we can plot each frequency component separately. When the signal is incident from port 1 at $\omega_1$ (Fig. 4b), the field component at $\omega_1$ remains isolated on the first resonator and no field is excited *at frequency* $\omega_1$ in the second resonator, preventing transmission to port 2 at this frequency. This is consistent with the very small static coupling $k_0$ between the rings. However, a nonzero field exists in the second resonator at frequency $\omega_2$ (Fig. 4c), which is consistent with upward frequency conversion. This field leaks out to port 2, giving non-zero transmission. When the signal is incident from port 2 at $\omega_2$, the field at $\omega_2$ remains localized on resonator 2, and there is no significant transmission to port 1 (as shown in Fig. 4d). Due to the upward-only frequency conversion property, downward frequency conversion is not allowed and the Floquet harmonic at $\omega_1$ is identically zero (Fig. 4e), making the system an extremely efficient isolator. Naturally, the same phenomenon can happen in a system for which $\omega_1 > \omega_2$, in which case $\Omega$ needs to be negative, which reverses the helicity of the modulation and thus the direction of frequency conversion. Therefore, by cascading two inverted systems, one with upward frequency conversion, and one with downward frequency conversion, we can make a non-reciprocal amplifier that operates *without* changing the frequency of the signal. The main technical challenge is related to synchronizing the modulation signals.

In conclusion, we have shown that time-periodic modulation can be a direct source of non-Hermiticity in realistic physical systems, and that it can induce novel physical effects such as large non-reciprocal behavior, unidirectional parametric gain, and insertion-loss free isolation. Different from previous works about time-Floquet non-Hermitian systems [52-55], which considered the time-modulation of an already non-Hermitian static system, here non-Hermiticity is induced by the modulation itself. We have shown that large non-reciprocal parametric gain can be obtained by modulating the coupling term in such a way that its phase is linearly increasing with time, in order to provide upward-only frequency generation. The unidirectional amplification in our system is of parametric nature, and is obtained at a frequency much smaller than the signal frequency, different from usual parametric amplification that requires pumping at the double frequency (note that this property is also found in QASERs [56]). Our findings show that non-Hermitian time-Floquet systems can overcome the insertion loss challenges typical of Hermitian time-Floquet isolators, which may have practical implications by enabling full-duplex communication in the next generation of integrated telecommunication systems. A tunable microwave non-reciprocal amplifier may be readily implemented using voltage controlled capacitors and resistors combined with negative impedance converters [51], providing a new path to efficient signal isolation, multiplexing and demultiplexing. Our findings, based temporal coupled-mode theory, are very general and may be applied to various physical platforms, including microwaves, acoustics, and optical signals. Altogether, these exciting findings establish the richness of the physics of wave propagation in non-Hermitian time-dependent systems.

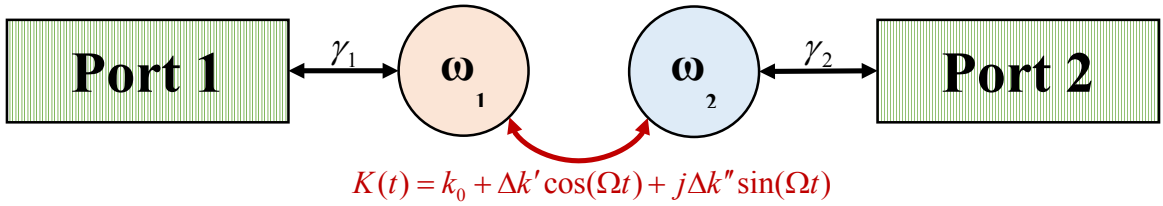

**Fig. 1: Non-Hermitian time-Floquet system under study.** We consider two coupled resonators with resonant frequencies $\omega_1$ and $\omega_2$, coupled to ports 1 and 2 with lifetimes $1/\gamma_1$ and $1/\gamma_2$, respectively. The two resonators are coupled together with a time-periodic coupling coefficient $K(t)$, with $K(t) = k_0 + \Delta k' \cos(\Omega t) + j\Delta k'' \sin(\Omega t)$.

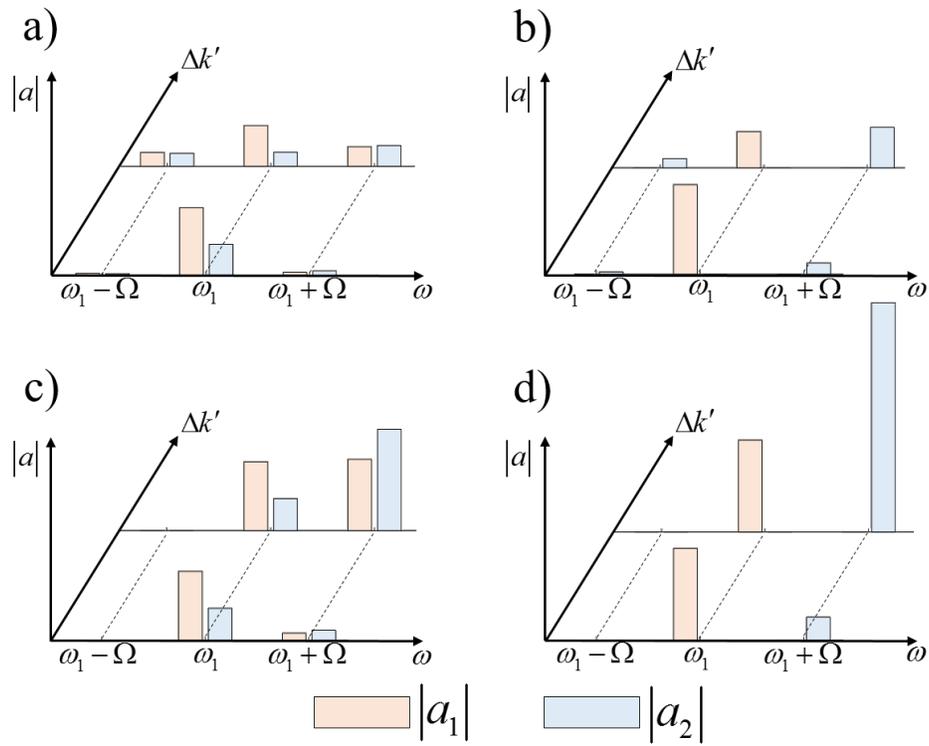

**Fig. 2: Excited Floquet amplitudes for excitation at $\omega_1$ from port 1.** The plots represent the amplitudes of the three dominant Floquet harmonics in the first (orange) and second (blue) resonators for two different (arbitrary) values of $\Delta k'$, and for a) $\Delta k'' = 0$ and $k_0 \neq 0$, b) $\Delta k'' = 0$ and $k_0 = 0$, c) $\Delta k' = \Delta k''$ and $k_0 \neq 0$, d) $\Delta k' = \Delta k''$ and $k_0 = 0$. The system is excited at $\omega_1$ from port 1.

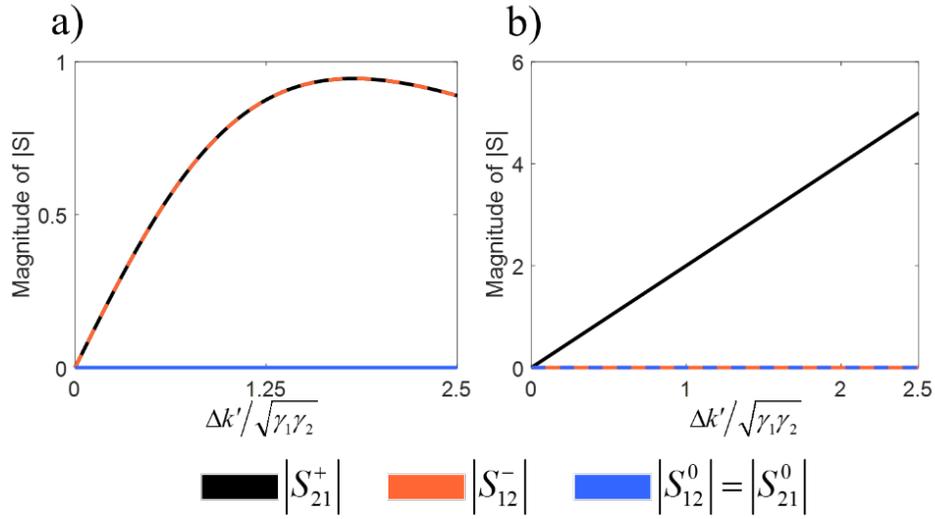

**Fig. 3: Non-reciprocal gain in non-Hermitian time-Floquet systems.** We plot the magnitude of the transmission coefficients to demonstrate the highly non-reciprocal behavior of the system. $S_{21}^0$ and $S_{21}^+$ correspond to the transmissions to port 2, respectively at $\omega_1$ and $\omega_2$, when a signal at $\omega_1$ is incident on port 1. $S_{12}^0$ and $S_{12}^-$ correspond to the transmissions to port 1, respectively at $\omega_2$ and $\omega_1$, when a signal at $\omega_2$ is incident on port 2. We compare the (reciprocal) Hermitian time-Floquet system with a) $\Delta k'' = 0$ and $k_0 = 0$ to the (non-reciprocal) non-Hermitian time-Floquet system b) $\Delta k' = \Delta k''$ and $k_0 = 0$. Transmission through any other frequency channel is identically zero.

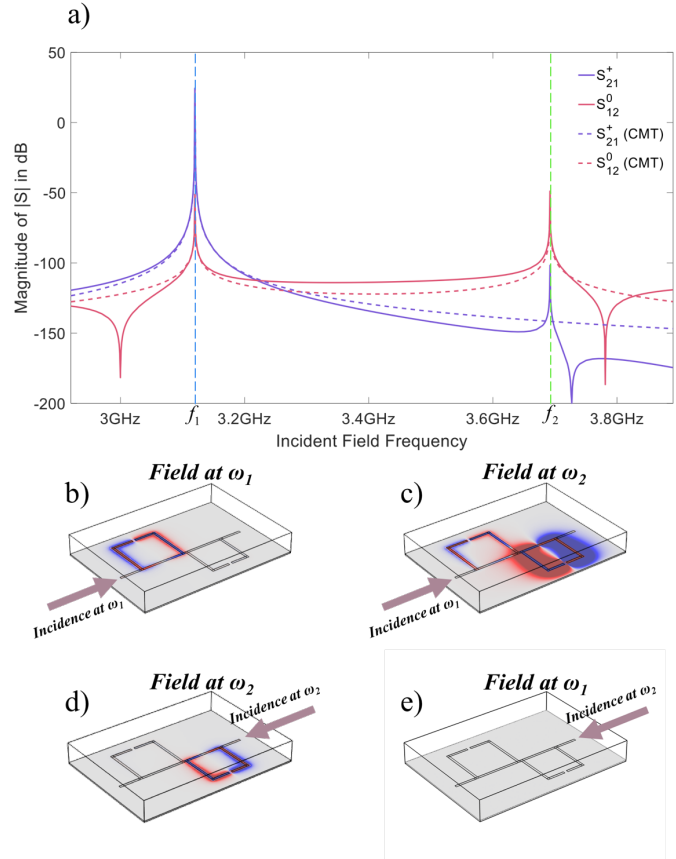

**Fig. 4: Full-wave finite-element demonstration of non-reciprocal gain at microwave frequencies.** The system is made of two microstrip ring resonators connected with a time-modulated capacitor in series with a time-modulated resistor [51]. a) Spectrum of the scattering parameters demonstrating strong non-reciprocity and one-way amplification in a non-Hermitian time-Floquet system made of coupled split-ring resonators. At $f_1$, $|S_{21}^+|$ is more than 30 dB, and the difference between $|S_{21}^+|$ at $f_1$ and $|S_{12}^0|$ at any frequency is more than 70 dB. Panels b) and c) correspond to incidence on port 1 at $\omega_1$ and show the vertical electric field component at $\omega_1$ and $\omega_2$, respectively. Panels d) and e) correspond to incidence on port 2 at $\omega_2$ and show the vertical electric field components at $\omega_2$ and $\omega_1$, respectively. All panels are plotted with the same scale, where blue correspond to a negative field and red to a positive field.